\documentclass[aps,pra,reprint,footinbib]{revtex4-2}
\usepackage{graphicx}
\usepackage{amsmath}
\usepackage[T1]{fontenc}
\usepackage[utf8]{inputenc}
\usepackage{hyperref}

\begin{document}

\title{Statistical properties of cold bosons in a ring trap}
\date{\today}
\author{Maciej Kruk}
\affiliation{Center for Theoretical Physics, Polish Academy of Sciences, Aleja Lotników 32/46, 02-668 Warsaw, Poland}
\author{Maciej Łebek}
\affiliation{Center for Theoretical Physics, Polish Academy of Sciences, Aleja Lotników 32/46, 02-668 Warsaw, Poland}
\author{Kazimierz Rzążewski}
\affiliation{Center for Theoretical Physics, Polish Academy of Sciences, Aleja Lotników 32/46, 02-668 Warsaw, Poland}

\begin{abstract}
A study of an interacting system of bosons in a ring trap at a finite temperature is presented. We consider a gas with contact and long-range dipolar interactions within a framework of the classical fields approximation. For a repulsive gas we have obtained coherence length, population of the ground state and its fluctuations as a function of temperature. In the case of an attractive gas we study local density fluctuations.  Additionally, we exactly calculate the partition function for the ideal gas in the canonical ensemble and derive several other macroscopic state functions.
\end{abstract}

\maketitle

\section{Introduction}\label{intro}
    The physics of Bose-Einstein condensation has been intensively studied since the appearance of the first papers published in the 1920s~\cite{Bose1924,*Einstein1924,*Einstein1925}. Theoretical research investigated not only population of the condensate, but also its fluctuations. The problem of the latter turned out to be more troublesome. E. Schr{\"o}dinger was the first to observe that the standard theory of noninteracting gas in the grand canonical ensemble predicts unphysically large fluctuations~\cite{schrodinger1989statistical}. Later on, it was noted that different fluctuations are found in different statistical ensembles~\cite{Ziff1977a}.
    
    The new era of BEC studies began in 1995, when the rubidium and sodium gases were condensed in laboratories~\cite{Davis1995a,*Anderson1995a}. Not surprisingly, that achievement accelerated the progress of theory. Papers concerning canonical~\cite{Politzer1996a} and microcanonical~\cite{Navez1997a} ensemble fluctuations in the experimentally relevant case of the harmonic trap appeared shortly after. These results were followed by studies of interacting gases in the later years ~\cite{Bienias2011c,Bienias2011b,Bhattacharyya2016a,Idziaszek1999a}. It is worth stressing that there are still many open questions in this area, e.g. results are inconclusive, for details see~\cite{Kristensen2019a}.
    
    Over ten years ago, local density fluctuations of quasi-1D Bose gas were measured ~\cite{Esteve2006a}. However, until now, there were no experimental data concerning fluctuations of the population of the condensate. The recent pioneering measurements of fluctuations ~\cite{Kristensen2019a} revive the interest in such problem in the case of weakly interacting Bose gas.
    
    In this paper, we present detailed studies of statistical properties of the interacting Bose gas confined to a toroidal trap. Such a confinement makes the system effectively one dimensional.
     A trap of that form has been succesfully realized in experiment ~\cite{Meinert2015a}.
     
    We analyze two types of interaction between particles: contact and dipolar long-range ones as defined in~\cite{Sinha2007a}.
    For dipolar interaction, both attractive and repulsive character can be achieved. In case of the repulsive interactions, we study number of atoms in the condensate and its fluctuations as a function of temperature. Additionally, we analyze the impact of temperature on a coherence length. For an attractive case, i.e. in the presence of bright solitons, we investigate local density fluctuations. Results concerning interacting system are supported by exact analytic expressions for ideal gas obtained in the canonical ensemble. We consider system with number of particles equal $N=100$. 
     
    Paper is structured in the following way.  In Sec. \ref{model} we present the most important theoretical aspects of our model that are common to all problems that we adress and discuss our main methods using which we obtain macroscopic state functions of the gas. Sec. \ref{repulsive} and Sec. \ref{attractive} are dedicated exclusively to theoretical extensions of our model and results concerning repulsive and attractive gas. Paper is ended with a brief summary in Sec. \ref{summary} containing the most important conclusions of our work. In the Appendix we present analytic formulas for partition function, correlation function and fluctuations of the zero momentum component for the ideal gas in the canonical ensemble.
\section{The model}\label{model}
    The Hamiltonian of the system we consider in this paper reads
    \begin{equation}
    \label{eq:hamiltonian}
    \begin{split}
        \hat{H} &= \int \mathrm{d}x \; {\hat{\psi}}^{\dag}(x) \frac{\hat{p}^2}{2m} \hat{\psi}(x) + \\
        & + \int \int \mathrm{d}x \, \mathrm{d}x' \; {\hat{\psi}}^{\dag}(x) {\hat{\psi}}^{\dag}(x') \hat{V}(x-x')\hat{\psi}(x') \hat{\psi}(x)
    \end{split}
    \end{equation}
    with $\hat{p}$ - momentum operator, $m$ - mass of the particle and $\hat{V}(x-x')$ - interaction potential.
    
    In order to cope with the difficult problem of interacting particles, we employ the classical fields approximation. It consists of replacing the atomic field operator $\hat{\psi}(x)$ by a complex c-number function $\psi(x)$ (for details see~\cite{Brewczyk2007a}). We can tune it to give us correct macroscopic state functions, provided we choose the optimal momentum cutoff.
    The use of such an approximation is analogous to using Maxwell equations instead of QED in the case of light, for which a cutoff is necessary to avoid the UV catastrophe known in the theory of black-body radiation.
    
    The main idea behind the classical fields approximation is to replace creation and annihilation operators by complex amplitudes and to neglect modes with value of momentum higher than cutoff momentum $k_{max}$. The problem of choosing the proper value of the cutoff is discussed in detail in repulsive and attractive gas sections. 
    
    In this paper we rely on mathematical equivalence between particles on the ring and free particles with periodic boundary conditions with period equal to the length of the trap $L$. The atomic field within the classical fields approximation
    \begin{equation}
        \psi (x) = \sum _{-k_{max}} ^{k_{max}} \alpha  _k  \, \varphi _k (x)=\sum _{-k_{max}} ^{k_{max}}  \frac{1}{\sqrt{L}} \, \alpha _k \, e^{ikx},
    \end{equation}
    where the set $\varphi _k$ is the orthonormal basis in the single-particle Hilbert space and $k=\frac{2 \pi n}{L},  \,n=0,\pm 1,... , \pm n_{max}$. Energy corresponding to state $\varphi _k$ is equal $E_k=\frac{\hbar ^2 k^2}{2m}=\frac{2 \pi ^2 \hbar ^2}{mL^2}n^2$. From now we will use $\epsilon \equiv \frac{2 \pi ^2 \hbar ^2}{m L^2}$, $\hbar/ \epsilon$ and $L$ as units of energy, time and length respectively. Hamiltonian as a function of the complex amplitudes $\alpha_j$ reads
    \begin{equation}
    \label{Halpha}
        H=\sum_{-n_{max}}^{n_{max}}j^2|\alpha _j|^2+\frac{1}{2} \sum_{j_1,j_2,j_3,j_4}C_{j_1,j_2,j_3,j_4} \, \alpha^*_{j_1}\alpha^*_{j_2}\alpha_{j_3}\alpha_{j_4},
    \end{equation}
    where
    \begin{equation}
    \begin{aligned}
        &C_{j_1,j_2,j_3,j_4}= \\
        &=\int_0^{1}\int_0^{1}\mathrm{d}x  \, \mathrm{d}x' \; e^{-2 \pi i[(j_1-j_3)x+(j_2-j_4)x']} \;V(x-x').
        \end{aligned}
    \end{equation}
    Additionally, we obtain set of equations of motion for amplitudes $\alpha _j$. Plugging Hamiltonian \eqref{eq:hamiltonian} into the Heisenberg equation and replacing operators with complex amplitudes yields
    \begin{equation}
    \label{eq:eqs}
        i  \, \frac{\mathrm{d} \alpha_j}{\mathrm{d} t}=j^2 \alpha_j +\sum_{j_1,j_2,j_3}C_{j_1,j,j_2,j_3} \,\alpha^*_{j_1}\alpha_{j_2}\alpha_{j_3},
    \end{equation}
    where $j=0, \pm 1, \cdots ,\pm n_{max}$.
    
    We consider contact interparticle potential $V(x-x')=g \, \delta (x-x')$ and quasi-1D dipole-dipole interaction potential~\cite{Deuretzbacher2010a}
    \begin{equation}
    \begin{split}
        &V(x-x')=g \; \frac{1}{4l_{\perp}}\Bigg[-2\Big|\frac{x-x'}{l_{\perp}}\Big|+ \\ &+e^{\frac{1}{2}\big|\frac{x-x'}{l_{\perp}}\big|^2}\sqrt{2 \pi} \, \bigg(1+\Big|\frac{x-x'}{l_{\perp}}\Big|^2\bigg)\text{Erfc}\Bigg(\Big|\frac{x-x'}{\sqrt{2}l_{\perp}}\Big|\Bigg)\Bigg]
    \end{split}
    \end{equation}
    with dependence on $l_{\perp}=\sqrt{\frac{\hbar}{m \omega _{\perp}}}$, where $\omega _{\perp}$ is the frequency of harmonic trap responsible for transversal confinement of particles. Note that length $l_{\perp}$ is directly related to the width of the dipolar potential and the potential is normalized to the parameter $g$ quantifying the strength of the interactions~\cite{Deuretzbacher2010a}. In our model the parameter $g$ is determined by two factors: dipole moments and their orientation which is kept constant during motion of the particles. By proper orientation of dipoles on the ring we are able to obtain attractive ($g<0$) and repulsive ($g>0$) character of interactions. For $x \gg 0$ we observe $ V(x) \sim \frac{1}{x^3}$ as for the standard dipolar potential.
     
    The full formula for quasi-1D potential contains additional $\delta$-interaction term. However, we want to focus on differences between short and long-range interactions, that is why we consider the situation where this term can be neutralized. 
    One can achieve this by proper tuning of Feshbach resonances leading to the cancellation of $\delta$ terms.
    
    Due to the ring geometry or equivalently, periodic boundary conditions, the expression for long-range potential should be modified 
     \begin{equation}
         V_{per}(x-x')=\sum_{n=-\infty}^{\infty} V(x-x'-n).
     \end{equation}

    In this paper we will use canonical ensemble for a system described with the classical fields approximation. To obtain estimates of statistical properties, we sample the ensemble with a Monte Carlo algorithm.
     
    First, to get the states from the thermal equilibrium distribution of the canonical ensemble we decided to use Metropolis algorithm \cite{Metropolis1953a} implemented as described in \cite{Witkowska2010a} with code available here \footnote{\url{github.com/mbkruk/BoseGas}}. We compare these results to those  obtained from the time evolution of equations \eqref{eq:eqs}.
    
    The equations of motion look similar to those studied by Fermi, Pasta and Ulam ~\cite{Fermi1955}. For contact interactions they can be seen as set of equations describing evolution of Fourier amplitudes of function $\psi(x)$ satisfying periodic nonlinear Schr{\"o}dinger equation (NLS). It is known that periodic NLS have infinite number of constants of motion ~\cite{faddeev2007hamiltonian}. It means that in principle our system is not ergodic. Three constants of motion have clear interpretation of energy, momentum and the number of particles. That is why we can think of time-averaged quantities as quantities obtained in a microcanonical ensemble further constrained by the constant momentum and the remaining constants of motion. Note that for the ideal gas ($g=0$) populations of modes do not change in time, hence the equations \eqref{eq:eqs} are not adequate to describe time evolution of noninteracting system.

\section{The repulsive gas}\label{repulsive}
    As indicated in the introduction, this section covers two independent subjects: population of the condensate together with its fluctuations and coherence length. Each one of these two problems needs a different choice of the cutoff parameter. That aspect of classical fields approximation - the need to change the cutoff to obtain correct values of different state functions, was discussed in detail in the paper \cite{Pietraszewicz2018a}.  In both situations our point of reference to establish criterion for cutoff are exact results for the ideal gas in the canonical ensemble.
\subsection{Population of the condensate and its fluctuations}
    We start this section with discussion of the cutoff parameter $k_{max}$ which plays an important role in our calculations. To determine its optimal value, we use similar approach to ~\cite{Witkowska2009a,Witkowska2010a}. We turn to noninteracting gas of $N$ bosons and calculate probability distribution $P(N_{ex})$  of having $N_{ex}$ atoms excited \eqref{eq:Pexact}. We use the canonical ensemble and compare our results to similar distribution $P_{cl}(N_{ex})$ which has the same physical meaning but is obtained using classical fields approximation \eqref{eq:Pclass}. For more details see the Appendix, where we derive formulas \eqref{eq:Pexact} and \eqref{eq:Pclass}. We introduce standard notation $\beta =1/k_B T$.

    \begin{widetext}
    \begin{equation}
    \label{eq:Pexact}
     P(N_{ex})=\frac{\sum_{j=1}^{\infty} e^{-\beta E_j N_{ex}} \prod_{\substack{k=1\\k \neq j}}^{\infty} \frac{1}{(1-e^{-\beta(E_k-E_j)})^2}\Bigg(N_{ex}+1 + 2\sum_{\substack{l=1\\l \neq j}}^{\infty} \frac{1}{1-e^{-\beta(E_j-E_l)}} \Bigg)}{\prod _{j=1}^{\infty} \frac{1}{(1-e^{-\beta E_j})^2}-\sum_{j=1}^{\infty}\frac{e^{-\beta E_j N}}{e^{ \beta E_j}-1}\prod_{\substack{k=1\\k \neq j}}^{\infty} \frac{1}{(1-e^{-\beta(E_k-E_j)})^2} \Bigg( N+1 +\frac{1}{1-e^{-\beta E_j}}+2\sum_{\substack{l=1\\l \neq j}}^{\infty} \frac{1}{1-e^{-\beta(E_j-E_l)}} \Bigg)}
    \end{equation}
    
    \begin{equation}
    \label{eq:Pclass}
     P_{cl}(N_{ex})=\frac{\sum_{j=1}^n e^{-\beta E_j N_{ex}} \prod_{\substack{k=1\\k \neq j}}^n \frac{1}{(\beta E_k -\beta E_j)^2} \Bigg(N_{ex}+2\sum_{\substack{l=1\\l \neq j}}^n \frac{1}{\beta E_j-\beta E_l}\Bigg)}{\prod_{j=1}^n \frac{1}{(\beta E_j)^2}\\-\sum_{j=1}^n  \frac{e^{-\beta E_j N}}{\beta E_j} \prod_{\substack{k=1\\k \neq j}}^n \frac{1}{(\beta E_k -\beta E_j)^2}\Bigg(N+\frac{1}{\beta E_j}+2\sum_{\substack{l=1\\l \neq j}}^n \frac{1}{\beta E_j-\beta E_l}\Bigg) }.
    \end{equation}
    \end{widetext}
    
\begin{figure}[h]
    \includegraphics{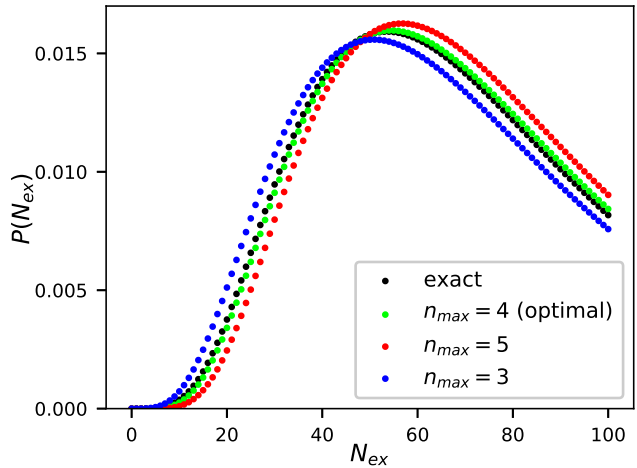}
    \caption{Comparison of exact  distribution $P(N_{ex})$ and discretized classical fields distribution $P_{cl}(N_{ex})$ obtained for different values of cutoff $n_{max}$. The temperature equals $\beta =0.03625$. We observe that cutoff obtained from relation \eqref{eq:cutoff} provides the best agreement between curves.}
    \label{fig:comp}
    \end{figure}

    Continous distribution \eqref{eq:Pclass} is discretized. We do that by dividing the interval $[0,N]$ into $N+1$ subintervals of equal size and averaging the distribution over subsequent subintervals.
    Optimal cutoff corresponds to the situation, where two distributions:  exact and discretized classical match each other ( see Fig. \ref{fig:comp}).
    
    Comparing $P(N_{ex})$ and $P_{cl}(N_{ex})$ leads us to the conclusion that cutoff parameter should be determined from following relation
    \begin{equation}
    \label{eq:cutoff}
     n_{max} = \sqrt{C k_B T}
    \end{equation}
    with $C=0.58$. 

\begin{figure*}[!]
    \includegraphics{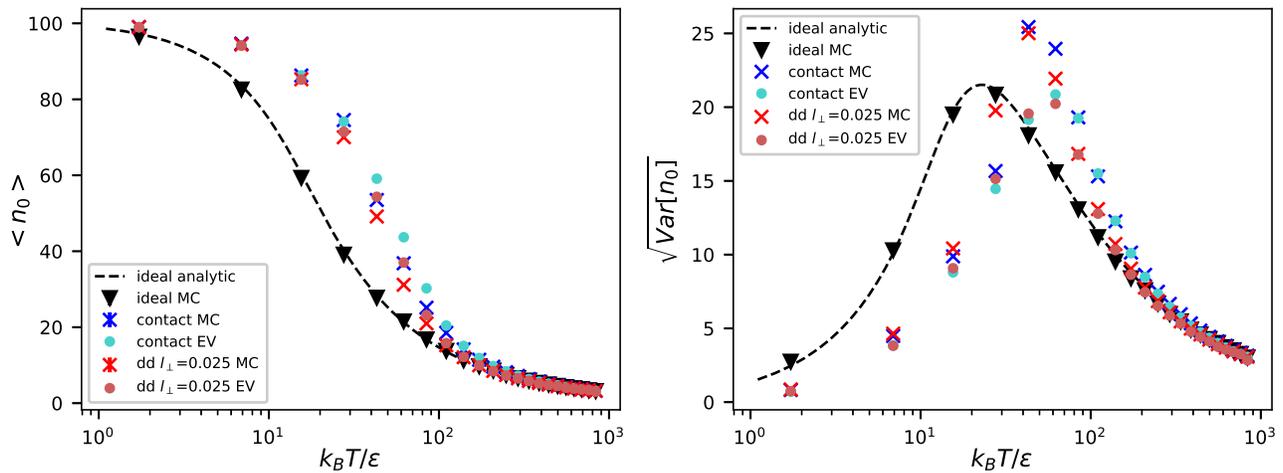}
    \caption{Average occupation of the condensate (left) and its fluctuations (right) as a function of temperature. We have obtained results for ideal and interacting gas with $g=1.0$. For the ideal gas we are able to reproduce with MC simulations exact analytic results calculated in canonical ensemble. We compare contact interactions and dipolar with $l_{\perp}=0.025$. Note that fluctuations obtained in time evolution are smaller than those from MC simulations. Contrary to 1D harmonic trap \cite{Bienias2011b}, depletion with temperature of interacting gas is slower than in case of the ideal gas.}
    \label{fig:stat}
    \end{figure*}

    With a suitable cutoff we are able to obtain average number of atoms in condensate and its fluctuations as a function of temperature. We use two approaches described in Sec. \ref{model}. Our results are presented in Fig. \ref{fig:stat}.
    
    Initial conditions for the evolution were chosen from the set of states generated by the Monte Carlo algorithm in such a way  that the energy of the state was the closest to the equilibrium value and the momentum was closest to zero. Two states randomly selected in this way in principle  may differ in values of other constants of motion, which in turn may lead to somewhat different values of time-averaged quantities. However, we have observed that for sufficiently strong interactions time averaged fluctuations and populations do not depend on the choice of initial condintions provided the energies and momenta are the same.
    
    Microcanonical ensemble, as well as our equations, do not involve temperature - the temperature is not a control parameter of microcanonical ensemble. Thus, the temperature in the case of time evolution should not be seen as a physical parameter, but rather as an indicator telling us that the evolution was performed for the  occupation of modes and energy characteristic for a system in equilibrium at the temperature $T$.
    
    The interaction strength used for simulations was $g=1.0$. In order to check whether this value corresponds to the regime of weakly interacting gas, we have calculated quantum depletion of the condensate at $T=0$ within the standard Bogoliubov approximation. The depletion turned out to be around 14 percent.

\subsection{Coherence length}
    The problem of phase coherence and closely related momentum distribution of the gas in 1D harmonic trap was investigated both experimentally \cite{Richard2003a} and theoretically \cite{Kadio2005a}.

    Due to dimensionality of our system, for high enough temperatures, we enter the regime where so-called quasicondensation occurs. Quasicondensation refers to the situation where there is no single dominant eigenvalue of single-particle density matrix. In other words, there is more than one mode with macroscopic population. This phenomenon is clearly visible in exact results concerning the ideal gas and occurs for repulsive gas as well. In this paragraph we look at the normalized correlation function (brackets $\langle \cdot \rangle$ denote canonical ensemble average)
    \begin{equation}
    \label{eq:correlation}
     g_1(x-x')=\frac{\langle \hat{\psi}^{\dag}(x)\hat{\psi}(x')\rangle}{\sqrt{\langle|\hat{\psi}(x')|^2\rangle \langle|\hat{\psi}(x)|^2\rangle}}
    \end{equation}
    with a special attention given to coherence length measuring the rate of decay of the function \eqref{eq:correlation}. Correlation function depends heavily on occupations of higher energy modes, that is why, due to quasicondensation, we no longer can rely on criterion \eqref{eq:cutoff} which was established by matching distributions that focus on occupation of the ground state only and in a sense do not take into account the distribution of higher modes. Therefore, new formula for cutoff is needed.
    \begin{figure}[h]
    \includegraphics{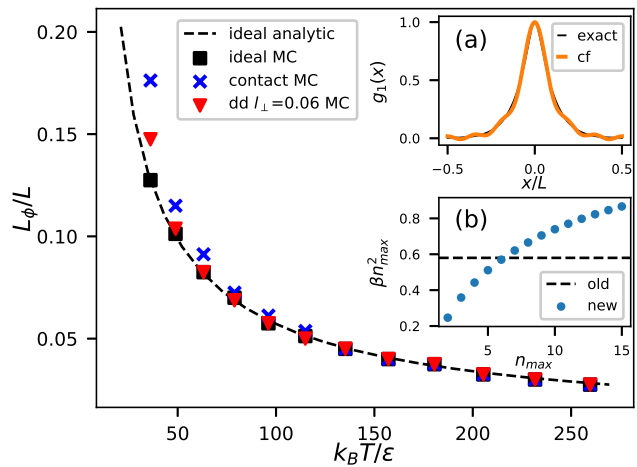}
    \caption{Coherence length as a function of temperature for $g=0.2$. We compare results for contact and dipolar interactions with $l_{\perp}=0.06$. With new values of cutoff parameter $n_{max}$ obtained by matching HWHM of exact and classical fields correlation function of the ideal gas $(a)$ we are able to reproduce with MC simulations coherence length calculated from analytic formulas for correlation function. The lower inset $(b)$ presents comparison of cutoff \eqref{eq:cutoff} and new cutoff suited for coherence length. From relation \eqref{eq:cutoff} we see that $\beta n_{max}^2=0.58=\text{const}$ contrary to values of $n_{max}$ obtained to match HWHMs.}
    \label{fig:coherence}
    \end{figure}

    Again, we consider noninteracting gas within the framework of canonical ensemble and calculate correlation function in exact treatment (with discrete occupations of states) and in classical fields approximation (for details see the Appendix). We have decided to measure the coherence length $L_{\phi}$ by half width at half maximum (HWHM) of function \eqref{eq:correlation}. Cutoff is chosen to match coherence length obtained from correlation function in exact treatment and in classical fields approximation (see Fig. \ref{fig:coherence}$(a)$). New values of cutoff parameter in relation to the old ones (from formula \eqref{eq:cutoff}) are presented in Fig. \ref{fig:coherence}$(b)$. We study the decay of coherence length with temperature (see Fig. \ref{fig:coherence}).
    It is interesting to notice that correlation function of the ideal gas  (see Fig. \ref{fig:coherence}$(a)$) is fairly smooth at the neighbourhood of zero and does not display cusp as it is in the case of gas in harmonic trap \cite{Kadio2005a}. The reason can be attributed to faster decay of occupations of subsequent energy levels that serve a role of coefficients in a expression for correlation function (for details see the Appendix). The difference is caused by different dependence of energy on quantum number which is linear for harmonic trap and quadratic in our system.
\section{The attractive gas}\label{attractive}
    Now we turn our attention to the attractive gas. This time even in low-temperature regime we expect quasicondensate. We observe that in results of Monte Carlo simulations, where even for very weak interactions and very low temperature, condensate population is far from 100 percent. Moreover, we expect bright solitons in desity profile of the gas.
\subsection{The cutoff}
    Because of quasicondensation for the attractive gas, we can no longer extend our ideal gas cutoff criterion \eqref{eq:cutoff}. The choice of the cutoff should be modified.
    
\begin{figure}
    \includegraphics{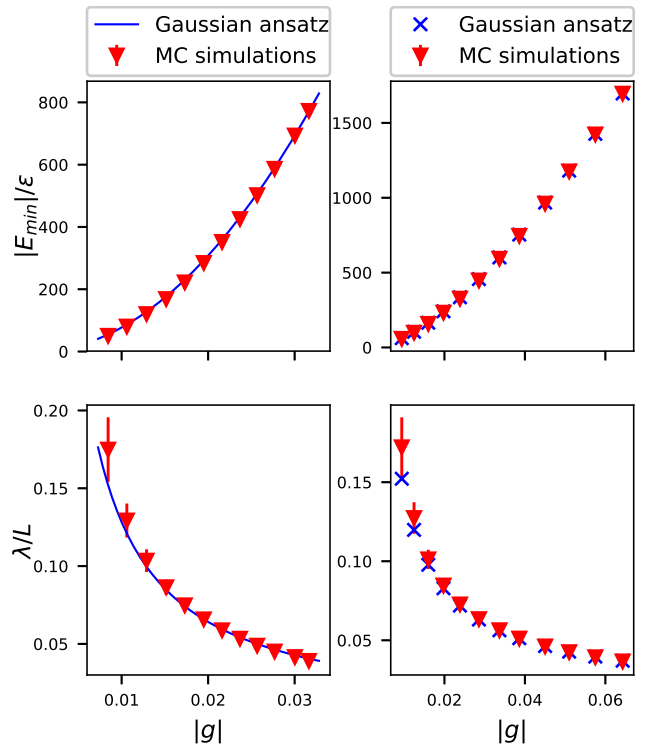}
    \caption{Results of reproducing the optimal state from Gaussian ansatz \eqref{eq:optimal} with MC simulations. The left panels present data for contact interactions, whereas the right panels correspond to the dipolar interaction with $l_{\perp}=0.025$. The first row compares average energy at very low temperature $\beta=0.58$ obtained using cutoff criterion \eqref{eq::cutoff2} with energy of optimal state with the same value of $g$. The values of interaction strength for which the MC simulations were performed correspond to the values of $n(g)=1,\ldots,12$. The second row presents comparison between the width of the optimal state and width from MC simulations. Note the differences for wide states with large $\lambda$.}
    \label{fig:reproducing}
    \end{figure}

    We start with finding an estimate of the ground state of attractive gas of $N$ particles. We employ two methods: analytical approximation and numerical algorithm. One can expect that the wave function of the ground state will take a localized shape due to the attractive interactions. Basing on this assumption we take the wave function of the following form
    \begin{equation}\label{eq:optimal}
     \Psi(x_1,...,x_N) =\prod _{j=1}^N \varphi(x_j), \hspace{0.16cm} \varphi(x)=\frac{1}{\sqrt{\sqrt{\pi}\lambda}} \text{e}^{-\frac{1}{2}(\frac{x}{\lambda})^2}
     \end{equation}
    with a free parameter $\lambda$ describing the width of the wave function. Such a form of the ansatz does not take the periodicity into account, therefore we expect it to describe our system well only if $\lambda$ is small enough. We then minimize the energy in this state with respect to $\lambda$. For contact interactions, the interaction energy can be computed analytically. The total energy reaches minimum for $\lambda =\frac{1}{ \sqrt{2} (\pi)^{3/2}|g| (N-1)} $ and equals
    \begin{equation}
     E_{\delta} = -\frac{1}{4}\pi g^2 (N-1)^2 N 
     \end{equation}
    However, it is not possible to analytically solve the dipole-dipole interaction and thus in this case the energy was computed numerically and the minimimum was found numerically. That is the reason why in Fig. \ref{fig:reproducing} for contact interactions we draw continous line, but for dipolar ones we have restricted ourselves to the values of interaction strength used to perform MC simulations.
   
    Our goal was to the modify cutoff in such a way that Monte Carlo simulations at very low temperature reproduce the energy and the width of the optimal state (see Fig. \ref{fig:reproducing}). This approach is similar to \cite{Bienias2011c} and can be expressed in formula
    \begin{equation}
    \label{eq::cutoff2}
        n_{max}=\sqrt{C k_B T}+n(g)
    \end{equation}
    where $n(g)$ is the number of mode pairs we have to add to reproduce the optimal state. We emphasize that extension of above formula to high temperatures is not obvious.
\subsection{Random walk}
    It is clear that the many-body Hamiltonian we consider commutes with the rotation generator. For this reason, the multi-particle wave function of the ground state should not change under rotations. However, when measuring positions of individual particles, one should rather expect localized Gaussian-like shape with well-defined position of maximum as a result. Of course, no point on the circle is distuinguished, so in the series of measurements, we should see uniform distribution of positions of the maximum.
    
    The phenomenon of spontaneous symmetry breaking via measurement was understood over twenty years ago in the case of interference of two condensates ~\cite{Javanainen1996a,Andrews1997a} and more recently, for excited state of 1D repulsive gas in the ring geometry ~\cite{Syrwid2015a}.
    
     \begin{figure}[h]
    \includegraphics{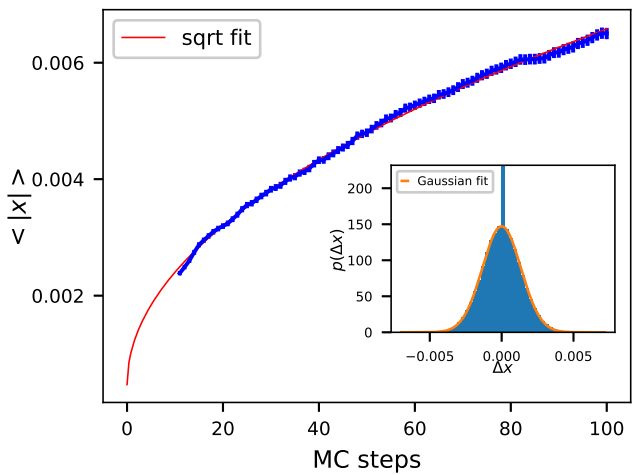}
    \caption{Average distance traveled by the wave packet maxima and a histogram of the shift in its position after a single MC step with corresponding function fits -- $\sqrt{n/a}+b$ and a normal distribution respectively. Note the spike in the histogram at $\Delta x=0$, the fit was done ignoring this spike. Interactions used for simulations were contact with interaction strength $g=-0.0152$. }
    \label{fig:randomwalk}
    \end{figure}
    
    We are able to somehow relate these remarks about breaking the rotational symmetry to results from MC simulations and time evolution. Firstly, we take the Metropolis algorithm and treat each consecutive sample as an element of a discrete time sequence. Then, we show that the peak of the wave profile exhibits Brownian motion properties (such as mean distance after $n$ time steps behaves like $\sqrt{n}$; each consecutive pair of changes in positions is not correlated). So, it is indeed Markovian random walk, despite the spike in the histogram presented in Fig. \ref{fig:randomwalk}. The spike is a delta-like addition to otherwise normal distrubution of position shifts. It is a byproduct of the Metropolis algorithm and more precisely, it is caused by the rejection rate of the new sets of alphas. The coefficient of the delta-like factor - the probability that there will be no change in a position caused by the rejection of a new set of alphas - directly corresponds and is equal to the rejection probability of the algorithm. By default, we optimize the rejection probability to be $0.5$ for fastest thermalization. The whole ensemble restores the rotational symmetry because samples do not distuinguish any point on the circle.
    It is interesting to note that within the classical fields approximation, individual realization of the field breaks symmetry and thus corresponds to a single measurement.
    
    Then, we provided the equations \eqref{eq:eqs} with initial conditions generated by MC algorithm for parameters guarrenteeing presence of bright solitons. The shape of the soliton was preserved during time evolution. We analyzed the positions of peak obtained in equal time steps. It turns out that the distribution of jump's length is Gaussian, but due to correlation between length of two consecutive jumps (peak tends to travel in one direction for quite a long time) we did not observe characteristic $\sqrt{t}$ behavior of a mean distance.
    \begin{figure}[h]
    \includegraphics{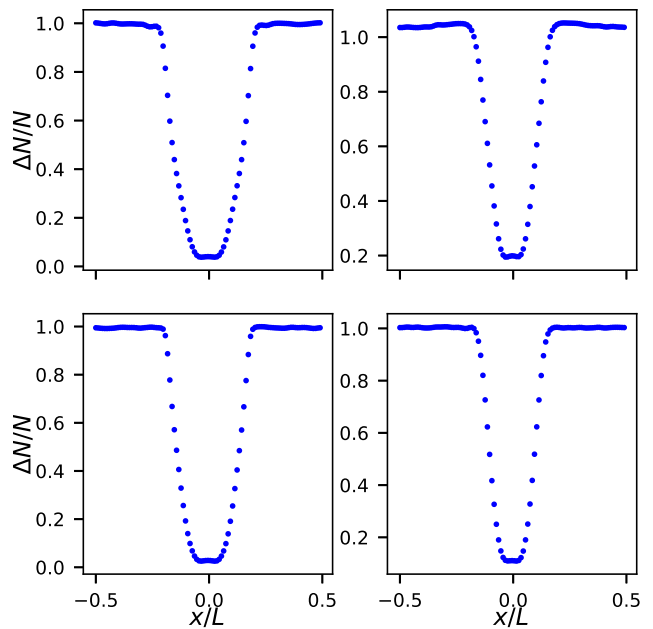}
    \caption{Local density fluctuations. The left panels present data for $\beta=0.58$, whereas for the right panels we have $\beta=0.0644$. We compare results for contact interaction (upper row) and dipolar with $l_{\perp}=0.06$ (lower row) with equal interaction strength $g=-0.0216$. The trap was divided into 101 bins.}
    \label{fig:localfluc}
    \end{figure}
\subsection{Local density fluctuations}
    The last issue of attractive gas is the problem of local density fluctuations. For every sample generated by Monte Carlo simulation, we consider its density profile.
    Every density profile corresponds to a different position of the soliton. That is why using bins with positions fixed relatively to the ring would not lead us to reveal any structure in the distribution of density fluctuations across bins. We use other approach in which we start with determining the position of the center of mass (CM) of our system. Then we divide the ring into bins with respect to the position of the CM. Finally, we collect data from all samples generated with MC algorithm and measure the variance of the number of atoms in each bin (see Fig. \ref{fig:localfluc}).
\section{Summary}\label{summary}
    We have thoroughly studied statistical properties of Bose gas confined to a ring trap, both for contact and dipolar interactions. Our results concern problems that are already accesible for experimental verification in harmonic traps. For repulsive gas, not only did we concentrated on average value and fluctuations of the zero momentum component, but also investigated the coherence length as a function of temperature, the quantity which depends on occupuation of higher energy modes. We have shown that with a proper criterion for the cutoff, classical fields approximation can tackle not only problems connected with occupation of the lowest energy mode, but can also be useful in the quasicondensation regime, in particular to study coherence length and local density fluctuations in the case of attractive gas.

    For gas with attractive interactions, apart from local density fluctuations we have analyzed the evolution of classical field throughout the Metropolis algorithm and the Heisenberg equation of motion. Our observations of the evolution of density profile lead us to conclusion that consecutive positions of the soliton display the same properties as one would expect from series of measurements of particles on the ring and therefore support interpretation that single realization of the classical field corresponds to a single measurement.

\begin{acknowledgments}
    We thank K. Pawłowski for enlightening discussions and P. Grochowski for careful reading of the manuscript. Authors were supported by (Polish) National Science Center Grant 2015/19/B/ST2/02820. Center for Theoretical Physics PAN is a member of KL FAMO.
\end{acknowledgments}

\appendix
\begin{widetext}
\section{Partition function of the ideal gas}
    We consider a system of $N$ particles on the ring in thermal contact with a heat reservoir at the temperature $T$. We start with calculating canonical partition function. By $n_i$ and $E_i$ we denote occupation and energy of $i$-th state. We introduce shift of the energy scale such that all energies are positive. Partition function takes the form
    \begin{equation}
     Z_N(\beta)=\sum_{n_{-\infty}=0}^{\infty} \ldots \sum_{n_{\infty}=0}^{\infty} e^{-\beta \sum_j E_j n_j} \; \delta_{N,\sum_j n_j}.
    \end{equation}
    We use  integral representation of the Kronecker delta
    \begin{equation}
     \delta_{N,\sum_j n_j}=\frac{1}{2\pi} \int_0^{2\pi} \mathrm{d} \xi \; e^{i \xi \big(N-\sum_j n_j\big)}.
    \end{equation}
    After changing order of summation and integration, we are able to evaluate sums. Note that in case of atoms on the ring all energies except $E_0$ are double-degenerated ($E_i=E_{-i}$). We arrive at
    \begin{equation}
     Z_N(\beta)=\frac{1}{2\pi} \int_0^{2\pi} \mathrm{d} \xi \; e^{i \xi N} \prod_{j=-\infty}^{\infty} \frac{1}{1-e^{-(i \xi + \beta E_j)}}=\frac{1}{2\pi} \int_0^{2\pi} \mathrm{d} \xi \; \frac{e^{i \xi N}}{1-e^{-(i\xi+\beta E_0)}}\prod_{j=1}^{\infty} \frac{1}{(1-e^{-(i \xi + \beta E_j)})^2}.
    \end{equation}
    In the next step we perform variable change $z=e^{i \xi}$ and end up with integrating function $f(z)=\frac{z^N}{z-e^{-\beta E_0}}\prod_{j=1}^{\infty}\frac{z^2}{(z-e^{-\beta E_j})^2}$ over the unit circle. We employ the residue theorem method. For $z=e^{-\beta E_0}$ there is a first-order pole and in $z=e^{-\beta E_k}$, $k \neq 0$ we have poles of order 2. We finally obtain partition function 
    \begin{equation}
    \label{eq:pfexact}
    \begin{split}
     Z_N(\beta)&=e^{-\beta E_0 N} \prod _{j=1}^{\infty} \frac{1}{(1-e^{-\beta(E_j-E_0)})^2}+\\
     &+\sum_{j=1}^{\infty}\frac{e^{-\beta E_j N}}{1-e^{-\beta(E_0-E_j)}}\prod_{\substack{k=1\\k \neq j}}^{\infty} \frac{1}{(1-e^{-\beta(E_k-E_j)})^2} \Bigg( N+1 +\frac{1}{1-e^{-\beta (E_j-E_0)}}+2\sum_{\substack{l=1\\l \neq j}}^{\infty} \frac{1}{1-e^{-\beta(E_j-E_l)}} \Bigg).
    \end{split}
    \end{equation}

\subsection*{Classical fields}
    Now we calculate partition function in the classical fields description of particles on the ring. It means that we have to introduce cutoff parameter $n$ and occupations of modes are no longer discrete $n_i=|\alpha _i |^2$. Partition function can be written in a form
    \begin{equation}
     Z_N^{cl}(\beta)=\int \frac{\mathrm{d}^2 \alpha_{-n}}{\pi} \ldots \int \frac{\mathrm{d}^2 \alpha_n}{\pi}  \; e^{- \beta \sum_j E_j |\alpha_j|^2}  \delta \bigg(N-\sum_{j=-n}^n |\alpha_j|^2 \bigg).
    \end{equation}
    We use  the Dirac delta representation
    \begin{equation}
     \delta \bigg(N-\sum_{j=-n}^n |\alpha_j|^2 \bigg)=\frac{1}{2 \pi} \int_{-\infty}^{\infty} \mathrm{d} \xi \; e^{i \xi \big(N-\sum_j |\alpha_j|^2\big)}.
    \end{equation}
    Hence
    \begin{equation}
     Z_N^{cl}(\beta)=\frac{1}{2\pi}\int_{-\infty}^{\infty} \mathrm{d} \xi \; e^{i \xi N} \cdot {\Large I_{-n} \ldots I_n}, \qquad I_k=\int \frac{\mathrm{d}^2 \alpha_{k}}{\pi}e^{-|\alpha_k|^2(\beta E_k+i \xi)}=\frac{1}{\beta E_k+ i \xi}.
    \end{equation}
    All energies except $E_0$ are double-degenerated ($E_i=E_{-i}$)
    \begin{equation}
    \label{eq:pfclint}
     Z_N^{cl}(\beta)=\frac{1}{2 \pi} \int_{-\infty}^{\infty} \mathrm{d} \xi \; e^{i \xi N} \prod _{j=-n}^{n} \frac{1}{\beta E_j+i \xi}=\frac{(-1)^n}{2\pi i}\int_{-\infty}^{\infty} \mathrm{d} \xi \; e^{i \xi N} \frac{1}{\xi- i\beta E_0} \prod_{j=1}^n\frac{1}{(\xi - i \beta E_j)^2}.
    \end{equation}
    To calculate integral \eqref{eq:pfclint} we consider function $g(z):=\frac{e^{izN}}{z-i\beta E_0}\prod_{j=1}^n\frac{1}{(z-i\beta E_j)^2}$ and integrate it over semicircle in the upper half-plane using the residue theorem (integral over arc vanishes in the limit of infinite radius). For $z=i \beta E_0$ there is a pole of order 1, and for $z=i \beta E_k$ , $k \neq 0$ we have poles of order 2.
    Partition function follows from the residue theorem formula
    \begin{equation}
    \label{eq:pfclass}
    \begin{split}
     Z_N^{cl}(\beta)&=e^{-\beta E_0 N} \prod_{j=1}^n \frac{1}{(\beta E_j -\beta E_0)^2}+\\&+\sum_{j=1}^n \frac{e^{-\beta E_j N}}{\beta E_0-\beta E_j} \prod_{\substack{k=1\\k \neq j}}^n \frac{1}{(\beta E_k -\beta E_j)^2}\Bigg(N+\frac{1}{\beta E_j -\beta E_0}+2\sum_{\substack{l=1\\l \neq j}}^n \frac{1}{\beta E_j-\beta E_l}\Bigg).
    \end{split}  
    \end{equation}

\section{Distribution $P(N_{ex})$}
    We are interested in a probability distribution $P(N_{ex})$  of having $N_{ex}$ atoms excited and $N_0=N-N_{ex}$ occupying the ground state. To obtain the distibution we use simple formula ~\cite{Weiss1997a}
    \begin{equation}
    \label{eq:pnex}
     P(N_{ex})=\frac{Z_{N_{ex}}}{Z_N}.
    \end{equation}
    where $Z_{N_{ex}}$ is partition function calculated for situation where $N_{ex}$ atoms are excited and the rest occupy the ground state. We calculate $Z_{N{ex}}$ following the same steps as in the case of partition function \eqref{eq:pfexact}. Using formula \eqref{eq:pnex} and setting $E_0=0$ gives us 
    \begin{equation}
     P(N_{ex})=\frac{\sum_{j=1}^{\infty} e^{-\beta E_j N_{ex}} \prod_{\substack{k=1\\k \neq j}}^{\infty} \frac{1}{(1-e^{-\beta(E_k-E_j)})^2}\Bigg(N_{ex}+1 + 2\sum_{\substack{l=1\\l \neq j}}^{\infty} \frac{1}{1-e^{-\beta(E_j-E_l)}} \Bigg)}{\prod _{j=1}^{\infty} \frac{1}{(1-e^{-\beta E_j})^2}-\sum_{j=1}^{\infty}\frac{e^{-\beta E_j N}}{e^{ \beta E_j}-1}\prod_{\substack{k=1\\k \neq j}}^{\infty} \frac{1}{(1-e^{-\beta(E_k-E_j)})^2} \Bigg( N+1 +\frac{1}{1-e^{-\beta E_j}}+2\sum_{\substack{l=1\\l \neq j}}^{\infty} \frac{1}{1-e^{-\beta(E_j-E_l)}} \Bigg)}.
    \end{equation}
\subsection*{Classical fields}
    Again we use formula \eqref{eq:pnex} with $Z^{cl}_{N_{ex}}$ calculated similarly as \eqref{eq:pfclass} and obtain 
    \begin{equation}
     P_{cl}(N_{ex})=\frac{\sum_{j=1}^n e^{-\beta E_j N_{ex}} \prod_{\substack{k=1\\k \neq j}}^n \frac{1}{(\beta E_k -\beta E_j)^2} \Bigg(N_{ex}+2\sum_{\substack{l=1\\l \neq j}}^n \frac{1}{\beta E_j-\beta E_l}\Bigg)}{\prod_{j=1}^n \frac{1}{(\beta E_j)^2}\\-\sum_{j=1}^n  \frac{e^{-\beta E_j N}}{\beta E_j} \prod_{\substack{k=1\\k \neq j}}^n \frac{1}{(\beta E_k -\beta E_j)^2}\Bigg(N+\frac{1}{\beta E_j}+2\sum_{\substack{l=1\\l \neq j}}^n \frac{1}{\beta E_j-\beta E_l}\Bigg) }.
    \end{equation}
\section{Correlation function}
    In order to calculate coherence length we need to obtain  the function $\langle \hat{\psi}^{\dag}(x) \hat{\psi}(x') \rangle$, where brackets $\langle \cdot \rangle$ denote canonical ensemble average. We decompose atomic field in momentum basis
    \begin{equation}
    \label{eq:atomic}
     \hat{\psi}(x)= \sum _q \frac{1}{\sqrt{L}} \, \hat{a}_q \, e^{iqx}
    \end{equation}
    and write
    \begin{equation}
     \langle \hat{\psi}^{\dag}(x) \hat{\psi}(x') \rangle=\frac{1}{L}\sum_{q,q'}\ \langle \hat{a}_q^{\dag} \hat{a}_{q'}\rangle \,  e^{i(q'x'-qx)}=\frac{1}{L}\sum_{q} \langle n_q \rangle  \, e^{i q(x'-x)} = \frac{1}{L}\bigg(\langle n_0 \rangle +2\sum_{q=1}^{\infty} \langle n_q \rangle \, \cos \big(q(x-x')\big) \bigg).
    \end{equation}
    Above we used the fact that $\langle \hat{a}_q^{\dag} \hat{a}_{q'}\rangle = \delta _{q,q'} \langle n_q \rangle$ and $\langle n_q \rangle = \langle n_{-q} \rangle$. As we see, the whole calculation boils down to obtaining average occupation of all modes. This can be done in a similar way as the partition function (using the residue theorem) or simply by differentiating expression for partition function \eqref{eq:pfexact}.
    For $q=0$ we have
    \begin{equation}
    \label{eq:avzero}
     \langle n_0 \rangle= \frac{1}{Z_N(\beta)}\sum_{j=0}^{\infty} e^{-\beta E_j (N-1)}\prod_{\substack{k=0 \\ k \neq j}}^{\infty} \frac{1}{(1-e^{-\beta(E_k-E_j)})^2}\Bigg(N+2\sum_{\substack{l=0 \\ l \neq j}}^{\infty}\frac{1}{1-e^{-\beta(E_j-E_l)}} \Bigg),
    \end{equation}
    whereas for $q \neq 0$ we obtain
    \begin{equation}
    \begin{split}
     \langle n_q \rangle &= \frac{1}{Z_N(\beta)}\Bigg[ \frac{e^{-\beta E_q}}{(1-e^{-\beta E_q})^3} \prod_{\substack{j=1 \\ j \neq q}}^{\infty}\frac{1}{(1-e^{-\beta E_j})^2}-\sum_{\substack{j=1 \\ j \neq q}}^{\infty} \frac{e^{-\beta E_j N} e^{-\beta (E_q-E_j)}}{(e^{\beta E_j}-1)(1-e^{-\beta (E_q-E_j)})} \prod_{\substack{k=1 \\ k \neq j}}^{\infty}\frac{1}{(1-e^{-\beta(E_k-E_j)})^2} \times \\
     & \times \Bigg( N+ \frac{1}{1-e^{-\beta(E_j-E_q)}}+ \frac{1}{1-e^{-\beta E_j}} +2 \sum_{\substack{l=1 \\ l \neq j}}^{\infty} \frac{1}{1-e^{-\beta(E_j-E_l)}}\Bigg)-\frac{e^{-\beta E_q N}}{e^{\beta E_q}-1} \prod_{\substack{j=1 \\ j \neq q}}^{\infty}\frac{1}{(1-e^{-\beta(E_j-E_q)})^2} \times \\ 
     & \times \Bigg( \frac{N(N+1)}{2}+ \frac{N}{(1-e^{-\beta E_q})^2}+2\Big( N+1+\frac{1}{1-e^{-\beta E_q}}\Big)\sum_{\substack{k=1 \\ k \neq q}}^{\infty} \frac{1}{1-e^{-\beta (E_q-E_k)}} + \sum_{\substack{k=0 \\ k \neq q}}^{\infty} \frac{1}{(1-e^{-\beta(E_q-E_k)})^2}+ \\
     &+ 2\sum _{\substack{k,k'=1 \\ k,k' \neq q \\ k \neq k'}}^{\infty} \frac{1}{(1-e^{-\beta(E_q-E_k)})(1-e^{-\beta(E_q-E_{k'})})}+\frac{1}{2}\sum_{\substack{k=1 \\ k \neq q}}^{\infty} \frac{1}{\text{sh}^2\big(\frac{1}{2}\beta(E_q-E_k)\big)}-\frac{N}{4}\frac{1}{\text{sh}^2\big(\frac{1}{2}\beta E_q\big)} \Bigg) \Bigg].
    \end{split}
    \end{equation}
\subsection*{Classical fields}
    Replacing annihilation operators with complex amplitudes in expression \eqref{eq:atomic} leads to classical version of atomic field. From that follows the correlation function
    \begin{equation}
     \langle \psi^{*}(x) \psi(x') \rangle=\frac{1}{L}\sum_{q,q'}\ \langle \alpha_q^* \alpha_{q'}\rangle \,  e^{i(q'x'-qx)}=\frac{1}{L}\sum_{q} \langle |\alpha_q|^2 \rangle  \, e^{i q(x'-x)} = \frac{1}{L}\bigg(\langle |\alpha_0|^2 \rangle +2\sum_{q=1}^{\infty} \langle |\alpha_q|^2 \rangle \, \cos \big(q(x-x')\big) \bigg).
    \end{equation}
    Using similar methods as in the case of partition function (the residue theorem) or by differentiation of \eqref{eq:pfclass} we obtain for $q=0$
    \begin{equation}
     \langle |\alpha_0|^2 \rangle = \frac{1}{Z_N^{cl}(\beta)}\sum_{j=0}^n e^{-\beta E_j N} \prod_{\substack{k=0 \\ k \neq j}}^n \frac{1}{(\beta E_k - \beta E_j )^2} \Bigg(N+2\sum_{\substack{l=0 \\ l \neq j}}^n \frac{1}{\beta E_j-\beta E_l} \Bigg),
    \end{equation}
    for $q \neq 0$ we have
    \begin{equation}
    \begin{split}
     \langle |\alpha _q|^2 \rangle &=\frac{1}{Z_N^{cl}(\beta)}\Bigg[\frac{1}{\beta E_q} \prod_{j=1}^n \frac{1}{(\beta E_j)^2}- \sum_{\substack{j=1 \\ j\neq q}}^n e^{-\beta E_j N} \frac{1}{\beta E_j}\frac{1}{\beta E_q-\beta E_j} \prod_{\substack{k=1 \\ k \neq j}}^n\frac{1}{(\beta E_k -\beta E_j)^2} \times \\
     & \times\Bigg(N+\frac{1}{\beta E_j -\beta E_q}+\frac{1}{\beta E_j}+2 \sum_{\substack{l=1 \\ l \neq j}}^n \frac{1}{\beta E_j -\beta E_l} \Bigg)- e^{-\beta E_q N}\frac{1}{\beta E_q} \prod_{\substack{j=1 \\ j \neq q}}^n \frac{1}{(\beta E_q-\beta E_j)^2} \times \\
     & \times \Bigg(\frac{N^2}{2}+\frac{N}{\beta E_q}+ \frac{1}{(\beta E_q)^2}+2\Big(N+\frac{1}{\beta E_q}\Big) \sum_{\substack{k=1 \\ k \neq q}}^n \frac{1}{\beta E_q -\beta E_k}+\\
     &+2\sum_{\substack{k,k'=1 \\ k,k' \neq q \\ k \neq k'}}^n \frac{1}{(\beta E_q- \beta E_k)(\beta E_q- \beta E_{k'})}+3\sum_{\substack{k=1 \\ k \neq q}}^n \frac{1}{(\beta E_q-\beta E_k)^2} \Bigg) \Bigg].
    \end{split}
    \end{equation}
\section{Fluctuations}
    The variance of population of the condensate is equal $\text{Var}(n_0)=\langle n_0^2 \rangle -\langle n_0 \rangle ^2$. We have already found average population \eqref{eq:avzero}, the remaining component is
    \begin{equation}
    \begin{split}
     \langle n_0 ^2 \rangle &=\frac{1}{Z_N(\beta)}\Bigg[ \prod_{j=1}^{\infty} \frac{1}{(1-e^{-\beta E_j})^2}\Bigg(N^2-2(2N+1)\sum_{k=1}^{\infty}\frac{1}{e^{\beta E_k}-1}+4\sum_{\substack{k,k'=1 \\k \neq k'}}^{\infty}\frac{1}{(e^{\beta E_k}-1)(e^{\beta E_{k'}}-1)}+ \\
     &+2\sum_{k=1}^{\infty}\frac{1}{(e^{\beta E_k}-1)^2}+\sum_{k=1}^{\infty}\frac{1}{\text{sh}^2(\frac{1}{2}\beta E_k)} \Bigg)-\sum_{j=1}^{\infty}\frac{e^{-\beta E_j (N-1)}}{(e^{\beta E_j}-1)^3} \prod_{\substack{k=1 \\ k \neq j}}^{\infty}\frac{1}{(1-e^{-\beta(E_k-E_j)})^2} \times \\
     & \times\Bigg(1+\Big(e^{\beta E_j}+1\Big)\bigg(N+\frac{1}{e^{\beta E_j}-1}+2\sum_{\substack{l=0 \\ l \neq j}}^{\infty}\frac{1}{1-e^{\beta(E_j-E_l)}} \bigg) \Bigg) \Bigg].
    \end{split}
    \end{equation}
\end{widetext}
\bibliography{ms}
\end{document}